\documentclass{iau}
\usepackage{graphicx}

\title[Constraining Stellar Winds of Young Sun-like Stars] %% give here short title %%
{Constraining Stellar Winds of Young Sun-like Stars}

\author[Colin P. Johnstone, Theresa L\"{u}ftinger, Manuel G\"{u}del, \& Bibiana Fichtinger]   %% give here short author list %%
{Colin P. Johnstone$^1$, Theresa L\"{u}ftinger$^1$, Manuel G\"{u}del$^1$, \and Bibiana Fichtinger$^1$}

\affiliation{$^1$ University of Vienna, Department of Astronomy, T\"{u}rkenschanzstrasse 17, 1180 Vienna, Austria}

\pubyear{2008}
\volume{xxx}  %% insert here IAU Symposium No.
\pagerange{119--126}
\jname{Title of your IAU Symposium}
\editors{A.C. Editor, B.D. Editor \& C.E. Editor, eds.}
\begin{document}

\maketitle

\begin{abstract}

As part of the project Pathways to Habitability (http://path.univie.ac.at/), we study the properties of the stellar winds of low-mass and Sun-like stars, and their influences on the atmospheres of potentially habitable planets.
For this purpose, we combine mapping of stellar magnetic fields with magnetohydrodynamic wind models.


\end{abstract}

\firstsection % if your document starts with a section,
\section{Introduction}

\begin{figure}[b]
\begin{center}
 \includegraphics[width=0.45\textwidth]{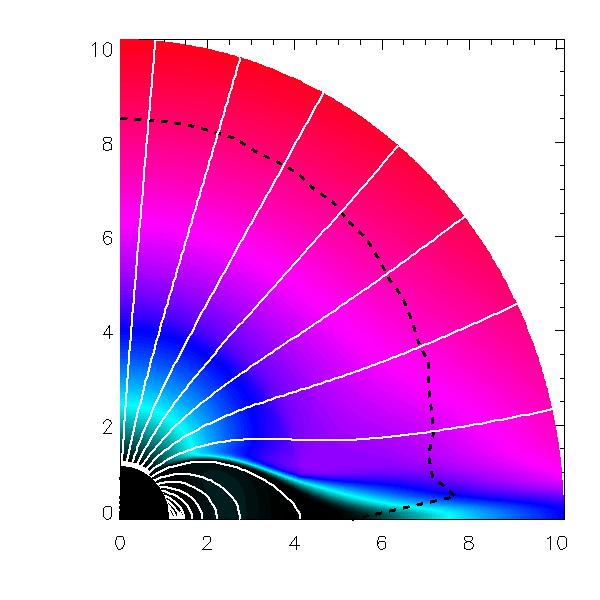} 
 \includegraphics[width=0.45\textwidth]{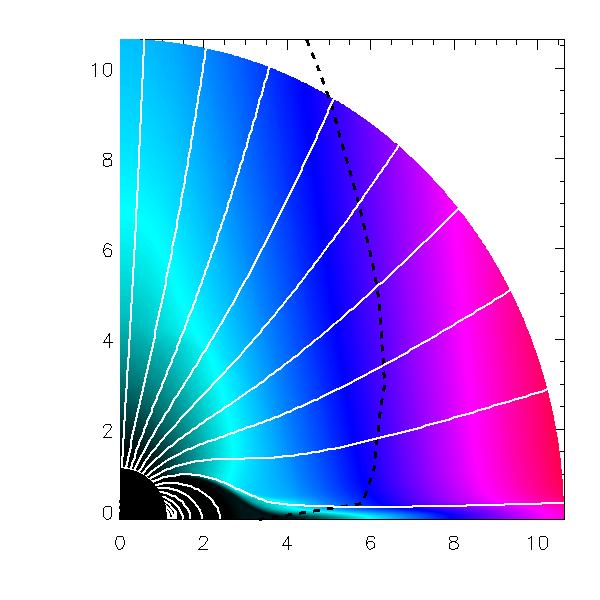} 
 \caption{
Velocity contour plots showing the expansion of 2D axisymmetric winds with a dipole magnetic field. 
These simulations were run using the MHD code Nirvana (\cite[Ziegler 2005]{Ziegler05}) for the cases of no stellar rotation (left) and a stellar rotation period of one day (right). 
The dashed black lines show the Alfv\'{e}n surfaces and the white lines show the structure of the magnetic field.
The black equatorial regions show regions of very low velocity, where the stellar plasma is contained within closed magnetic field lines.
 }
   \label{fig1}
\end{center}
\end{figure}

The Earth is embedded in the extended solar atmosphere we call the solar wind.
Such winds are known to emanate from other low-mass and Sun-like stars, including young solar analogues. 
Very little is currently known about how these winds influence the atmospheres of young habitable planets, largely because such winds are well constrained by neither observations nor theory.

The first physical model for the solar wind was developed by \cite[Parker (1958)]{Parker58}, who modelled the solar wind as a simple 1D isothermal pressure driven wind.
Later models of the solar wind produced used a polytropic equation of state, where the thermal pressure is given by $p \propto \rho^\gamma$, where $\gamma$ is the polytropic index (e.g. \cite[Totten et al. 1995]{Totten95}).
When $\gamma < 5/3$, these models implicitly heat the winds as they expand, but do not contain any description of the physics responsible for this heating. 
Despite a huge amount of progress observationally and theoretically, the fundamental mechanisms that drive the solar wind have not yet been determined. 

Although there have not been any direct detections of winds from low-mass or Sun-like stars, there are indirect methods of measuring these winds, most notably using measurements of Ly$\alpha$ absorption (\cite[Wood 2004]{Wood04}).
Attempts to directly detect thermal Bremsstrahlung radiation in radio from these winds have so far led to non-detections (\cite[Gaidos 2000]{Gaidos00}), putting important upper limits on the wind strengths.
As part of the project Pathways to Habitability, we attempt to detect the radio emission from these winds using VLT and ALMA observations.

\section{Wind Models}

We attempt to constrain the properties of stellar winds using magnetohydrodynamical modelling (e.g. see Fig. 1). 
We implicitly heat the winds by assuming a polytropic equation of state, as described above.
Similar models have been used successfully in previous studies (e.g. \cite[Keppens \& Goedbloed 1999]{Keppens99}; \cite[Matt \& Pudritz 2008]{Matt08}; \cite[Vidotto et al. 2009]{Vidotto09}).
Such models contain several free parameters, including the temperature and density at the base of the wind, and the polytropic index, $\gamma$.
Unfortunately, the free parameters are not well constrained and the results of these models (e.g. wind speeds and mass loss rates) are highly sensitive to these free parameters.

% Alternative models: Cranmer, Cohen, Suzuki

\section{Magnetic Field Observations}

Based on the Zeeman Doppler Imaging (ZDI) technique and the inversion of high-quality Stokes polarization data, we study the origin, strength, and distribution of magnetic fields on young active stars, including T Tauri stars, as a function of age, mass and spectral type. 
For modelling, we use a state-of-the-art ZDI code from \cite[Piskunov \& Kochukhov (2002)]{Piskunov02}, which has been extended for self-consistent temperature and magnetic mapping of cool active stars (\cite[Kochukhov \& Piskunov 2009]{Kochukhov09}) including treatment of molecular opacities. 
To obtain complete sets of circular and, for stars bright enough, linear Stokes parameter observations, we have applied and will apply for first-class spectropolarimetric data obtained with frontier instrumentation such as HARPSpol on the ESO 3.6m telescope, ESPaDOnS@CFHT (Hawaii), and NARVAL@TBL (France).


\begin{thebibliography}{}


\bibitem[Gaidos \etal\ (2000)]{Gaidos00}
{Gaidos, E. J., G\"{u}del, M., Blake, G. A.} 2000,
\textit{GeoRL}, 27, 501 

\bibitem[Keppens \& Goedbloed (1999)]{Keppens99}
{Keppens, R., \& Goedbloed, J. P.} 1999,
\textit{A\&A}, 343, 251 

\bibitem[Kochukhov \& Piskunov (2009)]{Kochukhov09}
{Kochukhov, O., \& Piskunov, N.} 2009,
\textit{ASPC}, 405, 539 

\bibitem[Matt \& Pudritz (2008)]{Matt08}
{Matt, S., \& Pudritz, R. E.} 2008,
\textit{ApJ}, 678, 1109

\bibitem[Parker (1958)]{Parker58}
{Parker, E.} 1958,
\textit{ApJ}, 128, 664 

\bibitem[Piskunov \& Kochukhov (2002)]{Piskunov02}
{Piskunov, N., \& Kochukhov, O.} 2002,
\textit{A\&A}, 381, 736 

\bibitem[Totten \etal\ (1958)]{Totten95}
{Totten, T. L., Freeman, J. W., \& Arya, S.} 1958,
\textit{JGR}, 100, 13 

\bibitem[Vidotto \etal\ (2009)]{Vidotto09}
{Vidotto, A. A., Opher, M., Jatenco-Pereira, V., \& Gambosi, T. I.} 2009,
\textit{ApJ}, 699, 441

\bibitem[Wood (2004)]{Wood04}
{Wood, B.} 2004,
\textit{LRSP}, 1, 2 

\bibitem[Ziegler (2005)]{Ziegler05}
{Ziegler, U.} 2005,
\textit{A\&A}, 435, 385 

\end{thebibliography}
\end{document}